\documentstyle[aps,preprint]{revtex}
\begin{document}
\newcommand{\pp}[1]{\phantom{#1}}
\newcommand{\be}{\begin{eqnarray}}
\newcommand{\ee}{\end{eqnarray}}
\newcommand{\ve}{\varepsilon}
\newcommand{\vs}{\varsigma}
\newcommand{\Tr}{{\rm Tr\,}}
\newcommand{\pol}{\frac{1}{2}}
\newtheorem{th}{Theorem}
\newtheorem{lem}[th]{Lemma}

\title{
Manifestly Covariant Approach to Bargmann-Wigner Fields (II):\\
From spin-frames to Bargmann-Wigner spinors
}
\author{Marek Czachor \cite{*}}
\address{
Wydzia{\l}  Fizyki Technicznej i Matematyki Stosowanej\\
 Politechnika Gda\'{n}ska,
ul. Narutowicza 11/12, 80-952 Gda\'{n}sk, Poland
}
\maketitle
\begin{abstract}
The Bargmann-Wigner (BW) scalar product is a particular
case of a larger class of scalar products parametrized by a
family of world-vectors. The choice of null and
$p$-dependent world-vectors leads to BW amplitudes 
which behave as local $SU(2)$ spinors (BW-spinors) 
if {\it passive\/}
transformations are concerned. 
The choice of null directions leads to a
simplified formalism which allows for an application of ordinary,
manifestly covariant spinor techniques in the context of
infinite dimensional unitary representations of the Poincar\'e
group. 
\end{abstract}

\section{Introduction}

In the first part of this work \cite{I} (henceforth called
``Part~I") we have introduced the most general form of
momentum-space 
Bargmann-Wigner (BW) scalar products \cite{BW}. 
The generalized products were
shown to depend on a family of world-vectors $\{t^a_k\}$ which
can be momentum-dependent. 

The standard form of the product introduced by Bargmann and
Wigner in \cite{BW}
corresponds to an (implicit) choice of a 
single future-pointing, timelike, and momentum-independent vector
$t^a$. This vector implicitly fixes a decomposition of the Minkowski
space into ``time" and ``space" which, accordingly, is used to
divide $SL(2,C)$ transformations into ``boosts" and ``rotations".
For fields defined on $m\neq 0$ mass shell
the ``rotations" possess finite dimensional representations
which are unitary  with respect to the
scalar product defined by $t^a$; 
``boosts" are represented unitarily by momentum-dependent
``rotations", the so-called Wigner rotations \cite{Ohnuki}. 
The momentum dependence of Wigner rotations makes the representation
infinite-dimensional. 

An assumption that generators of {\it unitary\/} representations of
symmetry groups are directly related to observable quantities 
is one of the most fundamental postulates of quantum theory.
On the other hand, {\it manifestly covariant\/} formulations of
classical relativistic theories proved
to be incredibly efficient from both physical and mathematical
point of view.  
The fact that the mentioned implicit decomposition into ``time" and
``space" is built so deeply into the structure of the unitary
representation makes it practically impossible to discuss
quantum theories in a manifestly covariant way directly at the
level of physical states.  In order to have covariant formulas 
in the standard approach
one has to switch to spinor wave functions which do not
have a direct probability interpretation. And {\it vice versa\/}:
If one wants to directly deal with probability amplitudes one
has to switch from spinors to BW amplitudes which are noncovariant.

The difficulties mentioned above motivated the author of this
paper to look for a
manifestly covariant reformulation
of the standard unitary representations of the Poincar\'e group.
In the next paper of this series we shall discuss the most
general form of generators obtained if one keeps the
world-vectors $\{t^a_k\}$ arbitrary. 

In this paper, however, we shall turn to another particular form
of unitary representations which seems to have been overlooked
until now: Those arising if one takes $t^a$ {\it null\/} and
{\it $p$-dependent\/}. The ``null formulation" is elegant and simple.
Its main technical advantage over the ``timelike" one is based
on the factorization property $t^{AA'}=\tau^A\bar \tau^{A'}$
typical of null world-vectors. This property leads directly from
spinors to BW wave functions, but the wave functions so obtained
are not equivalent to helicity amplitudes. The wave functions are
$SL(2,C)$ scalars if one considers {\it active\/} 
spinor transformations of the spinor BW fields, but become
$SU(2)$ spinors (BW-spinors) if one considers {\it passive\/}
transformations. 
BW-spinors can be treated by standard spinor methods. 
In addition to symplectic $\ve$-spinors used to
raise and lower BW-indices we introduce metric $\vs$-spinors 
used to express covariantly the BW scalar products. 

The proofs given below are based on particularly chosen
fields of spin-frames associated with 4-momenta. The explicit  
form of the spin-frames is given in Sec.~\ref{sec2}. In the rest
of the paper we do not make use of the explicit forms 
themselves but take advantage of some of their properties.

\section{Spin-frames associated with future pointing 4-momenta}
\label{sec2}
Let $S\in SL(2,C)$. The representations $({\pol},0)$, $(0,{\pol})$, 
$({\pol},0)\oplus (0,{\pol})$, and $(1,1)$ of $S$ will be denoted
by $S{_{A}}{^{B}}$, $\bar S{_{A'}}{^{B'}}$,
$S{_{\alpha}}{^{\beta}}$, and $\Lambda{_{a}}{^{b}}$,
respectively. We shall
occasionally skip the indices if no ambiguities arise. 
Spinor transformations of upper- and lower-index spinors are 
\be
T(S)\phi_{A}&=&S{_{A}}{^{B}}\phi_{B}\label{cov},\\
T(S)\phi^{A}&=&\phi^{B}S^{-1}{_{B}}{^{A}}=-S{^{A}}{_{B}}\phi^{B}.
\label{con}
\ee
Analogous transformations hold for primed spinors.
The convention differs slightly from this used in 
\cite{PR} (cf. Eq.~(3.6.1)). 

\subsection{Explicit spin-frame for $m\neq 0$}

Consider an arbitrary $p$-independent
spinor $\nu^A\neq 0$. 
Let $\omega^{a}=\omega^{A}\bar \omega^{A'}$, 
$\pi^a=\pi^A \bar \pi^{A'}$, where
\be
\omega^{A}
&=&\Bigl[\frac{m}{\sqrt{2}}\Bigr]^{1/2}\frac{\nu^A}{
\sqrt{p^{BB'}\nu_B
\bar \nu_{B'}}}=\omega^{A}(\nu,p)\label{oo}\\
\pi^A 
&=&\Bigl[\frac{\sqrt{2}}{m}\Bigr]^{1/2}
\frac{p^{AA'}\bar \nu_{A'}}
{\sqrt{p^{BB'}\nu_B
\bar \nu_{B'}}}=\pi^A(\nu, p).\label{pp}
\ee
Spinors (\ref{oo}), (\ref{pp}) satisfy 
\be
\omega^a\, p_a &=& m/\sqrt{2},\\
p^a &=&
\frac{m}{\sqrt{2}}\Bigl(\pi^a + \omega^a\Bigr),\\
\omega^A(\nu,p)\pi_A(\mu,p)&=&
\bar \omega^{A'}(\mu,p)\bar \pi_{A'}(\nu,p),\label{5}\\
\pi^A(\nu,p)\pi_A(\mu,p)&=&
\bar \omega^{A'}(\nu,p)\bar \omega_{A'}(\mu,p),\label{6}\\
\omega_A(\nu,p)\pi^A(\nu,p)&=&1,\label{s-f}\\
S{_{A}}{^{B}}\omega_{B}(\nu,\Lambda^{-1}p)&=&
\omega_{A}(S\nu,p),\label{Soo}\\
S{_{A}}{^{B}}\pi_{B}(\nu,\Lambda^{-1}p)&=&
\pi_{A}(S\nu,p),\label{Spp}
\ee
where $p^a$ is future-pointing and non-null, and $\nu_A$,
$\mu_A$ are arbitrary. 

\subsection{Explicit spin-frame for $m=0$}

Let $n^a$ be $p$-independent, timelike and future-pointing, and 
let $\nu_A\neq 0$ be also $p$-independent and arbitrary.
Define 
\be
\pi^A(\nu, p)
&=&
\frac{p^{AA'}\bar \nu_{A'}}
{\sqrt{p^{BB'}\nu_B
\bar \nu_{B'}}},\label{ppp}\\
\omega^{A}(\nu,n,p)
&=&
\frac{n^{AA'}\bar \pi_{A'}(\nu,p)}
{n^a\,p_a}=
\frac{n^{AA'}
p^{B}{_{A'}}\nu_{B}}
{n^a\,p_a\,\sqrt{p^{b}\,\nu_b
}},\label{ooo}
\ee
satisfying the spin-frame condition
\be
\pi_A(\nu, p)
\omega^{A}(\nu,p)=1
\ee
(note that the roles of $\pi$ and $\omega$ are reversed with
respect to the massive case --- this is consistent with the
notation from Part~I). Similarly to (\ref{Soo}), (\ref{Spp}) we
find 
\be
S{_{A}}{^{B}}\omega_{B}(\nu,n,\Lambda^{-1}p)&=&
\omega_{A}(S\nu,\Lambda n,p),\label{Soo'}\\
S{_{A}}{^{B}}\pi_{B}(\nu,\Lambda^{-1}p)&=&
\pi_{A}(S\nu,p).\label{Spp'}
\ee
An explicit calculation shows also that 
\be
\pi_A(\nu, p)\bar \pi_{A'}(\nu, p)=
\pi_A(\mu, p)\bar \pi_{A'}(\mu, p)=
p_{AA'},\label{pipi=p}
\ee
which implies
\be
\pi_A(\nu, p)=
\frac{p^{BB'}\mu_B\bar \nu_{B'}}
{|p^{CC'}\mu_C\bar \nu_{C'}|}
\pi_A(\mu, p).\label{fp}
\ee
Therefore two $\pi$-spinors having the same 
flagpole $p^a$ differ by
a phase.

\section{Passive transformations of BW
amplitudes (${m}\neq 0$)}

Let $S^a(p){_\alpha}{^\beta}$ denote the momentum-space
 Pauli-Lubanski (P-L)
vector for the bispinor ($m\neq 0$) representation and 
$S(\omega,p){_\alpha}{^\beta}=\omega^a S_a(p){_\alpha}{^\beta}$.

A Fourier transform of the Dirac bispinor expanded in
eigenstates of $S(\omega,p){_\alpha}{^\beta}$ is (cf. Eq.~(127)
in Part~I)
\be
\psi_\pm(\pm p)_\alpha=
\left(
\begin{array}{c}
\psi_\pm(\pm p)^{\it 0}_A\\
\psi_\pm(\pm p)^{\it 1}_{A'}
\end{array}
\right)=
-N\left(
\begin{array}{c}
\mp\omega_A(\nu,p)\,f^{(+)}_\pm(\nu,\pm p)+\pi_A(\nu,p)
\,f^{(-)}_\pm(\nu,\pm p)\\
\bar \pi_{A'}(\nu,p)\,f^{(+)}_\pm(\nu,\pm p)\pm\bar 
\omega_{A'}(\nu,p)\,f^{(-)}_\pm(\nu,\pm p)
\end{array}
\right)\label{bis}
\ee
where $N=\Bigl[\frac{m}{\sqrt{2}}\Bigr]^{1/2}$. The use of 
the italic font
distinguishes the ``BW-indices"
(``{\it 0\/}", ``{\it 1\/}") and
 the ordinary spinor ones (``0", ``1"). 
The amplitudes (cf. Appendix)
\be
N^{-1}\bar \omega^{A'}(\nu,p)\psi_\pm(\pm p)^{\it 1}_{A'}&=&
N^{-1}\omega^{\alpha}(\nu,p)\psi_\pm(\pm p)^{\it 1}_{\alpha}=
f^{(+)}_\pm(\nu,\pm p)=f^{\it 1}_\pm(\nu,\pm p),\\
N^{-1}\omega^{A}(\nu,p)\psi_\pm(\pm p)^{\it 0}_{A}&=&
N^{-1}\omega^{\alpha}(\nu,p)\psi_\pm(\pm p)^{\it 0}_{\alpha}=
f^{(-)}_\pm(\nu,\pm p)=f^{\it 0}_\pm(\nu,\pm p),\\
N^{-1}\omega^{A}(\nu,p)
\overline{\psi_\pm(\pm p)^{\it 1}_{A'}}&=&
N^{-1}\bar \omega^{\alpha'}(\nu,p)
\bar \psi_\pm(\pm p)^{\it 0}_{\alpha'}=
\overline{
f^{(+)}_\pm(\nu,\pm p)}=\bar f^{\it 0}_\pm(\nu,\pm p),\\
N^{-1}\bar \omega^{A'}(\nu,p)
\overline{\psi_\pm(\pm p)^{\it 0}_{A}}&=&
N^{-1}\bar \omega^{\alpha'}(\nu,p)
\bar \psi_\pm(\pm p)^{\it 1}_{\alpha'}=
\overline{
f^{(-)}_\pm(\nu,\pm p)}=\bar f^{\it 1}_\pm(\nu,\pm p)
\ee
are $SL(2,C)$ scalars. The $``(\pm)"$ indices refer to ``signs of
spin projections in null directions", so are not equivalent to
signs of helicity (which correspond to future timelike
directions). The signs $``\pm"$ written without braces
are signs of energy. 

A general BW field is a direct sum of inequivalent spinor
representations.
The numerical ($\it 0$ or $\it 1$) 
indices were introduced in Part~I to
distinguish between ``different components" of the field 
(that is those belonging to inequivalent representations
constituting the direct sum) and as such play a role of
a binary
numbering of the components.
The possibility of 
identifying $``+"$ with  ${\it 0}$ and 
$``-"$ with ${\it 1}$ is a particular property of the null
formalism introduced in Part~I. 

Spacetime traslations are represented in momentum representation
unitarily by 
one-dimensional phase factors. In the following sections we
shall concentrate only on the nontrivial part of the unitary
representation: the infinite dimensional representation of $SL(2,C)$.

Let $S\in SL(2,C)$.
An active bispinor transformation of the Dirac field
\be
\psi'_\pm(\pm p)_\alpha&=&
-N\left(
\begin{array}{c}
\mp\omega_A(\nu,p)\,{f'}^{\it 1}_\pm(\nu,\pm p)+\pi_A(\nu,p)
\,{f'}^{\it 0}_\pm(\nu,\pm p)\\
\bar \pi_{A'}(\nu,p)\,{f'}^{\it 1}_\pm(\nu,\pm p)\pm\bar 
\omega_{A'}(\nu,p)\,{f'}^{\it 0}_\pm(\nu,\pm p)
\end{array}
\right)\nonumber\\
&=&
S{_{\alpha}}{^{\beta}}\psi_\pm(\pm\Lambda^{-1}p)_\beta\nonumber\\
&=&
-N
\left(
\begin{array}{cc}
S{_{A}}{^{B}} & 0\\
0 & \bar S{_{A'}}{^{B'}}
\end{array}
\right)
\left(
\begin{array}{c}
\mp\omega_B(\nu,\Lambda^{-1}p)\,f^{\it 1}_\pm(\nu,\pm
\Lambda^{-1}p) +\pi_B(\nu,\Lambda^{-1}p)
\,f^{\it 0}_\pm(\nu,\pm \Lambda^{-1}p)\\
\bar \pi_{B'}(\nu,\Lambda^{-1}p)\,
f^{\it 1}_\pm(\nu,\pm \Lambda^{-1}p)\pm\bar 
\omega_{B'}(\nu,\Lambda^{-1}p)\,f^{\it 0}_\pm(\nu,\pm \Lambda^{-1}p)
\end{array}
\right)
\ee
induces a passive transformation of the amplitudes
\be
&\pp .&\left(
\begin{array}{c}
{f'}^{\it 0}_\pm(\nu,\pm p)\\
{f'}^{\it 1}_\pm(\nu,\pm p)
\end{array}
\right)\nonumber\\
&\pp .&\pp {=====}=
\left(
\begin{array}{rr}
-\omega^A(\nu,p)S{_{A}}{^{B}}\pi_B(\nu,\Lambda^{-1}p), &
\pm \omega^A(\nu,p)S{_{A}}{^{B}}\omega_B(\nu,\Lambda^{-1}p)\\
\mp\bar \omega^{A'}(\nu,p)\bar S{_{A'}}{^{B'}}\bar 
\omega_{B'}(\nu,\Lambda^{-1}p), &
-\bar \omega^{A'}(\nu,p)\bar S{_{A'}}{^{B'}}\bar 
\pi_{B'}(\nu,\Lambda^{-1}p)
\end{array}
\right)
\left(
\begin{array}{c}
{f}^{\it 0}_\pm(\nu,\pm \Lambda^{-1}p)\\
{f}^{\it 1}_\pm(\nu,\pm \Lambda^{-1}p)
\end{array}
\right).\label{t}
\ee
Using  (\ref{Soo}), (\ref{Spp}) 
we arrive at the following form of the transformation (\ref{t})
\be
\left(
\begin{array}{c}
{f'}^{\it 0}_\pm(\nu,\pm p)\\
{f'}^{\it 1}_\pm(\nu,\pm p)
\end{array}
\right)&=&
\left(
\begin{array}{rr}
\omega_A(\nu,p)\pi^A(S\nu,p), &
\mp \omega_A(\nu,p)\omega^A(S\nu,p)\\
\pm\bar \omega_{A'}(\nu,p)\bar 
\omega^{A'}(S\nu,p), &
\bar \omega_{A'}(\nu,p)\bar 
\pi^{A'}(S\nu,p)
\end{array}
\right)
\left(
\begin{array}{c}
{f}^{\it 0}_\pm(\nu,\pm \Lambda^{-1}p)\\
{f}^{\it 1}_\pm(\nu,\pm \Lambda^{-1}p)
\end{array}
\right).\label{t'}
\ee
Formula (\ref{t'}) shows that the ``BW-indices" play a dual 
role analogous to spinor indices if one considers
passive transformations of the BW amplitudes. To distinguish
between the ordinary spinor indices and the BW indices we shall
denote the latter by calligraphic letters. 
Therefore Eq.~(\ref{t'}) can be written in a compact form as 
\be
{f'}^{\cal A}_\pm(\nu,\pm p)=
{U(S)f}^{\cal A}_\pm(\nu,\pm p)=
{\cal U}_\pm(S,\nu,p){^{\cal A}}{_{\cal B}}
{f}^{\cal B}_\pm(\nu,\pm \Lambda^{-1}p).\label{t''}
\ee
Complex conjugated amplitudes 
$\bar f^{\it 0\/}=\overline{f^{\it 1\/}}$, 
$\bar f^{\it 1\/}=\overline{f^{\it 0\/}}$ transform according 
to 
\be
{\bar f}'^{\cal A}_\pm(\nu,\pm p)=
{\overline{U(S)f}}^{\cal A}_\pm(\nu,\pm p)=
\bar {{\cal U}}_\pm(S,\nu,p){^{\cal A}}{_{\cal B}}
{\bar f}^{\cal B}_\pm(\nu,\pm \Lambda^{-1}p).\label{bar-t''}
\ee

\section{Unitarity and unimodularity of BW transformation
matrices}

The matrices 
${\cal U}{^{\cal A}}{_{\cal B}}=
{\cal U}_\pm(S,\nu,p){^{\cal A}}{_{\cal B}}$ are
unimodular
\be
\ve_{\cal AB} &=& 
{\cal U}{^{\cal C}}{_{\cal A}}\,
{\cal U}{^{\cal D}}{_{\cal B}}\,
\ve_{\cal CD}\\
\ve^{\cal AB} &=& 
{\cal U}{^{\cal A}}{_{\cal C}}\,
{\cal U}{^{\cal B}}{_{\cal D}}\,
\ve^{\cal CD},
\ee
and unitary
\be
\vs_{\cal AB} &=&  
\bar {\cal U}{^{\cal C}}{_{\cal A}}\,
{\cal U}{^{\cal D}}{_{\cal B}}\,
\vs_{\cal CD}
\ee
(cf. Appendix).
To prove unimodularity we use properties 
(\ref{5}), (\ref{6}), (\ref{s-f}):
\be
\det {\cal U}
&=& 
\pol {\cal U}_{\cal AB}\,{\cal U}^{\cal AB}\nonumber\\
&=&
\omega_A(\nu,p)\pi^A(S\nu,p)
\bar \omega_{A'}(\nu,p)\bar \pi^{A'}(S\nu,p)
+
\omega_A(\nu,p)\omega^A(S\nu,p)
\bar \omega_{A'}(\nu,p)\bar \omega^{A'}(S\nu,p)\nonumber\\
&=&
\omega_A(\nu,p)\pi^A(S\nu,p)
\omega_{B}(S\nu,p)\pi^{B}(\nu,p)
+
\omega_A(\nu,p)\omega^A(S\nu,p)
\pi_{B}(\nu,p)\pi^{B}(S\nu,p)\nonumber\\
&=&
\omega_A(\nu,p)
\Bigl[
\omega_{B}(S\nu,p)\pi^A(S\nu,p)
-
\pi_{B}(S\nu,p)
\omega^A(S\nu,p)
\Bigr]
\pi^{B}(\nu,p)
=
\omega_A(\nu,p)
\pi^{A}(\nu,p)=1.
\ee
Unitarity follows from unimodularity and 
\be
{\cal U}{^{\it 0}}{_{\it 0}}&=&
\overline{
{\cal U}{^{\it 1}}{_{\it 1}}}=
\bar {\cal U}{^{\it 0}}{_{\it 0}},\\
{\cal U}{^{\it 1}}{_{\it 0}}&=&
-\,\overline{
{\cal U}{^{\it 0}}{_{\it 1}}}=-\,
\bar {\cal U}{^{\it 1}}{_{\it 0}}.
\ee
It is appropriate to recall the form of the scalar
product the representation $S\to U(S)$ is unitary with respect
to. Let $d\mu_m(p)$ denote an invariant measure on the mass-$m$
hyperboloid (cf. Eq.~(25) in Part~I). 
The scalar product is derived from the norm (130) in Part~I
which, using the $\vs_{\cal AB}$ BW-spinor, can be written as
\be
\parallel \psi_{\pm\alpha}\parallel'^2&=&
\int  d\mu_m(p)
\frac{
\omega^{A}(\nu,p)\bar \omega^{A'}(\nu,p)
\Bigl(
\psi{_\pm}(\pm p)^{\it 0}_A
\overline{\psi{_\pm}(\pm p)^{\it 0}_A}
+
\psi{_\pm}(\pm p)^{\it 1}_{A'}
\overline{\psi{_\pm}(\pm p)^{\it 1}_{A'}}
\Bigr) 
}
{\omega^a(\nu,p)\, p_a}\nonumber\\
&=&
\int d\mu_m(p)
N^{-2}\omega^{\alpha}(\nu,p)\bar \omega^{\alpha'}(\nu,p)
\psi{_\pm}(\pm p)^{\cal A}_\alpha
\bar \psi{_\pm}(\pm p)^{\cal B}_{\alpha'}
\vs_{\cal AB}\nonumber\\
&=&
\int d\mu_m(p)
f^{\cal A}_\pm(\pm p)
\bar f^{\cal B}_\pm(\pm p)
\vs_{\cal AB}
=
\int d\mu_m(p)
\Bigl(|f^{\it 0}_\pm(\pm p)|^2+|f^{\it 1}_\pm(\pm
p)|^2 \Bigr).
\label{exD}
\ee
The generalization to BW fields of arbitrary spin is immediate:
The norm (56) from Part~I becomes
\be
\parallel \psi_{\pm\alpha_1\dots\alpha_n}\parallel'^2&=&
\int d\mu_m(p)
f_\pm(\pm p)^{{\cal A}_1\dots {\cal A}_n}
\bar f_\pm(\pm p)^{{\cal B}_1\dots {\cal B}_n}
\vs_{{\cal A}_1{\cal B}_1}\dots
\vs_{{\cal A}_n{\cal B}_n}.
\label{n-exD}
\ee
The simplicity of formulas has been achieved because of the 
simultaneous use of $t^a=\omega^{A}(\nu,p)\bar
\omega^{A'}(\nu,p)$ in the generalized norm 
$\parallel\cdot\parallel'$, and in the expansion of the Dirac
bispinor in eigenstates of the P-L vector's projection in the
{\it same\/} $t^a$ direction.
The simplicity is lost if non-null $t^a$ are used since
only null world-vectors factorize. 
The privileged role played by the BW amplitudes obtained with
the help of the $p$-dependent spin-frames suggests that they deserve a
name of their own to distinguish them from the standard
``helicity" BW amplitudes. We will call the vectors
\be
f^{\cal A}=
\left(
\begin{array}{c}
f^{\it 0}\\
f^{\it 1}
\end{array}
\right)
\ee
the {\it BW-spinors\/}. BW-spinors can be also
regarded as $SU(2)$ spinor fields on a mass hyperboloid.

\section{Proof of $U(S')U(S)=U(S'S)$ ($m\neq 0$)}

Passive transformations (\ref{t''}) of the BW-spinors form a
unitary representation of the Poincar\'e group. The composition
property $U(S')U(S)=U(S'S)$ is proved by the following 
calculation:
\be
&\pp .&
\left(
\begin{array}{c}
{U(S')[U(S)f]}^{\it 0}_\pm(\nu,\pm p)\\
{U(S')[U(S)f]}^{\it 1}_\pm(\nu,\pm p)
\end{array}
\right)\nonumber\\
&\pp .&
\pp {xx}
=
\left(
\begin{array}{rr}
\omega_A(\nu,p)\pi^A(S'\nu,p), &
\mp \omega_A(\nu,p)\omega^A(S'\nu,p)\\
\pm\bar \omega_{A'}(\nu,p)\bar 
\omega^{A'}(S'\nu,p), &
\bar \omega_{A'}(\nu,p)\bar 
\pi^{A'}(S'\nu,p)
\end{array}
\right)\nonumber\\
&\pp .&\pp {xxxx}
\times
\left(
\begin{array}{rr}
\omega_B(\nu,\Lambda'{^{-1}}p)\pi^B(S\nu,\Lambda'{^{-1}}p), &
\mp \omega_B(\nu,\Lambda'{^{-1}}p)\omega^B(S\nu,\Lambda'{^{-1}}p)\\
\pm\bar \omega_{B'}(\nu,\Lambda'{^{-1}}p)\bar 
\omega^{B'}(S\nu,\Lambda'{^{-1}}p), &
\bar \omega_{B'}(\nu,\Lambda'{^{-1}}p)\bar 
\pi^{B'}(S\nu,\Lambda'{^{-1}}p)
\end{array}
\right)
\left(
\begin{array}{c}
{f}^{\it 0}_\pm(\nu,\pm (\Lambda'\Lambda)^{-1}p)\\
{f}^{\it 1}_\pm(\nu,\pm (\Lambda'\Lambda)^{-1}p)
\end{array}
\right)\nonumber\\
&\pp .&
\pp {xx} =
\left(              
\begin{array}{rr}
\omega_A(\nu,p)\pi^A(S'\nu,p), &
\mp \omega_A(\nu,p)\omega^A(S'\nu,p)\\
\pm\bar \omega_{A'}(\nu,p)\bar 
\omega^{A'}(S'\nu,p), &
\bar \omega_{A'}(\nu,p)\bar 
\pi^{A'}(S'\nu,p)
\end{array}
\right)\nonumber\\
&\pp .&\pp {xxxx}
\times
\left(
\begin{array}{rr}
\omega_B(S'\nu,p)\pi^B(S'S\nu,p), &
\mp \omega_B(S'\nu,p)\omega^B(S'S\nu,p)\\ 
\pm\bar \omega_{B'}(S'\nu,p)\bar 
\omega^{B'}(S'S\nu,p), &
\bar \omega_{B'}(S'\nu,p)\bar 
\pi^{B'}(S'S\nu,p)
\end{array}
\right)
\left(
\begin{array}{c}
{f}^{\it 0}_\pm(\nu,\pm (\Lambda'\Lambda)^{-1}p)\\
{f}^{\it 1}_\pm(\nu,\pm (\Lambda'\Lambda)^{-1}p)
\end{array}
\right)\nonumber\\
&\pp .&
\pp {xx} =
\left(
\begin{array}{rr}
\omega_A(\nu,p)\pi^A(S'S\nu,p), &
\mp \omega_A(\nu,p)\omega^A(S'S\nu,p)\\
\pm\bar \omega_{A'}(\nu,p)\bar 
\omega^{A'}(S'S\nu,p), &
\bar \omega_{A'}(\nu,p)\bar 
\pi^{A'}(S'S\nu,p)
\end{array}
\right)
\left(
\begin{array}{c}
{f}^{\it 0}_\pm(\nu,\pm (\Lambda'\Lambda)^{-1}p)\\
{f}^{\it 1}_\pm(\nu,\pm (\Lambda'\Lambda)^{-1}p)
\end{array}
\right)\nonumber\\
&\pp .&
\pp {xx} =
\left(
\begin{array}{c}
{U(S'S)f}^{\it 0}_\pm(\nu,\pm p)\\
{U(S'S)f}^{\it 1}_\pm(\nu,\pm p)
\end{array}
\right).
\nonumber
\ee
We have used here the following two sequences of identities:
\be
&\pp .&
\omega_A(\nu,p)\pi^A(S'\nu,p)
\omega_B(S'\nu,p)\pi^B(S'S\nu,p)
-
\omega_A(\nu,p)\omega^A(S'\nu,p)
\bar \omega_{B'}(S'\nu,p)\bar \omega^{B'}(S'S\nu,p)\nonumber\\
&\pp .&
=
\omega_A(\nu,p)\pi^A(S'\nu,p)
\omega_B(S'\nu,p)\pi^B(S'S\nu,p)
-
\omega_A(\nu,p)\omega^A(S'\nu,p)
\pi_{B}(S'\nu,p)\pi^{B}(S'S\nu,p)\nonumber\\
&\pp .&
=
\omega_A(\nu,p)
\Bigl[
\omega_B(S'\nu,p)
\pi^A(S'\nu,p)
-
\pi_{B}(S'\nu,p)
\omega^A(S'\nu,p)
\Bigr]
\pi^{B}(S'S\nu,p)\nonumber\\
&\pp .&
=
\omega_A(\nu,p)
\pi^{A}(S'S\nu,p),
\ee
and
\be
&\pp .&
\omega_A(\nu,p)\pi^A(S'\nu,p)
\omega_B(S'\nu,p)\omega^B(S'S\nu,p)
+
\omega_A(\nu,p)\omega^A(S'\nu,p)
\bar \omega_{B'}(S'\nu,p)\bar \pi^{B'}(S'S\nu,p)
\nonumber\\
&\pp .&=
\omega_A(\nu,p)\pi^A(S'\nu,p)
\omega_B(S'\nu,p)\omega^B(S'S\nu,p)
+
\omega_A(\nu,p)\omega^A(S'\nu,p)
\omega_{B}(S'S\nu,p)\pi^{B}(S'\nu,p)
\nonumber\\
&\pp .&=
\omega_A(\nu,p)\omega_B(S'\nu,p)
\pi^A(S'\nu,p)\omega^B(S'S\nu,p)
-
\omega_A(\nu,p)\pi_{B}(S'\nu,p)
\omega^A(S'\nu,p)\omega^{B}(S'S\nu,p)
\nonumber\\	
&\pp .&=
\omega_A(\nu,p)
\Bigl[
\omega_B(S'\nu,p)
\pi^A(S'\nu,p)
-
\pi_{B}(S'\nu,p)
\omega^A(S'\nu,p)
\Bigr]
\omega^{B}(S'S\nu,p)
\nonumber\\
&\pp .&=
\omega_A(\nu,p)
\omega^{A}(S'S\nu,p).
\ee

\section{Transformation properties of amplitudes for $m=0$}

The Hertz-type form of solutions of the massless BW equations
discussed in Part~I leads to a single BW amplitude $f_\pm(\pm p)$
for a massless field whose spin is arbitrary. 
This fact agrees with the general theorem
stating that a massless finite-spin irreducible 
unitary representation of the Poincar\'e group must be
induced by a one-dimensional representation.

An active transformation of the massless spinor field 
induces a passive transformation of the amplitude:
\be
\psi'_\pm(\pm p)^{{\it 0}\dots {\it 0}}_{A_1\dots A_n}&=&
\pi_{ A_1}(\nu,p)\dots \pi_{ A_n}(\nu,p)
f'_\pm(\nu,n,\pm p)^{{\it 0}\dots {\it 0}}\nonumber\\
&=&
S{_{A_1}}{^{B_1}}\dots
S{_{A_n}}{^{B_n}}
\pi_{ B_1}(\nu,\Lambda^{-1}p)\dots \pi_{ B_n}(\nu,\Lambda^{-1}p)
f_\pm(\nu,n,\pm \Lambda^{-1}p)^{{\it 0}\dots {\it 0}}
.\label{twist}
\ee
The passive transformation of the amplitude is (compare (\ref{fp}))
\be
U(S)f_\pm(\nu,n,\pm p)^{{\it 0}\dots {\it 0}}&=&
\Bigl[
\omega^{A}(\nu,n,p)\pi_{A}(S\nu,p)
\Bigr]^n
f_\pm(\nu,n,\pm \Lambda^{-1}p)^{{\it 0}\dots {\it 0}}\nonumber\\
&=&
\Biggl[
\frac{p^{AA'}\nu_A\bar S{_{A'}}{^{B'}}\bar \nu_{B'}}
{|p^{CC'}\nu_C\bar S{_{C'}}{^{D'}}\bar \nu_{D'}|}
\Biggr]^n
f_\pm(\nu,n,\pm \Lambda^{-1}p)^{{\it 0}\dots {\it 0}}
\label{phasefactor}\\
&=&
{\cal U}(S,\nu,p)
f_\pm(\nu,n,\pm \Lambda^{-1}p)^{{\it 0}\dots {\it 0}}.
\ee
(\ref{phasefactor}) shows that 
${\cal U}(S,\nu,p)$ is a phase factor and hence the
transformation is unitary.

Had we started with a massless field having $n$ primed indices
\be
\psi_\pm(\pm p)_{A'_1\dots A'_n}^{{\it 1}\dots {\it 1}}=
\bar \pi_{ A'_1}(\nu,p)
\dots\bar \pi_{ A'_n}(\nu,p)
f_\pm(\nu,n,\pm p)^{{\it 1}\dots
{\it 1}}, 
\ee
we would have obtained a complex-conjugated transformation rule
\be
U(S)f_\pm(\nu,n,\pm p)^{{\it 1}\dots {\it 1}}=
\bar {\cal U}(S,\nu,p)
f_\pm(\nu,n,\pm \Lambda^{-1}p)^{{\it 1}\dots {\it 1}}.
\ee
It is interesting that the sign-of-energy index ``$\pm$" is not
necessary in either ${\cal U}(S,\nu,p)$ or $\bar {\cal
U}(S,\nu,p)$. In the $m\neq 0$ case these signs entered the
transformation properties {\it via\/} the off-diagonal elements
of the $SU(2)$ matrices. Here the off-diagonal elements do not
appear since the representation is one-dimensional. 

\section{Proof of $U(S')U(S)=U(S'S)$ ($m=0$)}

It is sufficient to prove the composition property 
$U(S')U(S)=U(S'S)$ for a ``{\it 0\/}$\dots${\it 0\/}" amplitude:
\be
&\pp .&
U(S')[U(S)f]_\pm(\nu,n,\pm p)^{{\it 0}\dots {\it 0}}
\nonumber\\
&\pp .&\pp {xxxxx}=
{\cal U}(S',\nu,p){\cal U}(S,\nu,\Lambda'^{-1}p)
f_\pm(\nu,n,\pm (\Lambda'\Lambda)^{-1}p)^{{\it 0}\dots {\it 0}}
\nonumber\\
&\pp .&\pp {xxxxx}=
\Bigl[
\omega^{A}(\nu,n,p)\pi_{A}(S'\nu,p)
\omega^{B}(\nu,n,\Lambda'^{-1}p)\pi_{B}(S\nu,\Lambda'^{-1}p)
\Bigr]^n
f_\pm(\nu,n,\pm (\Lambda'\Lambda)^{-1}p)^{{\it 0}\dots {\it 0}}
\nonumber\\
&\pp .&\pp {xxxxx}=
\Bigl[
\omega^{A}(\nu,n,p)\pi_{A}(S'\nu,p)
\omega^{B}(S'\nu,S'n,p)\pi_{B}(S'S\nu,p)
\Bigr]^n
f_\pm(\nu,n,\pm (\Lambda'\Lambda)^{-1}p)^{{\it 0}\dots {\it 0}}
\nonumber\\
&\pp .&\pp {xxxxx}=
\Bigl[
\omega^{A}(\nu,n,p)
\Bigl\{
\ve_{A}{^B}
+
\omega_{A}(S'\nu,S'n,p)
\pi^{B}(S'\nu,p)
\Bigr\}
\pi_{B}(S'S\nu,p)
\Bigr]^n
f_\pm(\nu,n,\pm (\Lambda'\Lambda)^{-1}p)^{{\it 0}\dots {\it 0}}
\nonumber\\
&\pp .&\pp {xxxxx}=
\Bigl[
\omega^{A}(\nu,n,p)
\pi_{A}(S'S\nu,p)
\Bigr]^n
f_\pm(\nu,n,\pm (\Lambda'\Lambda)^{-1}p)^{{\it 0}\dots {\it 0}}
\nonumber\\
&\pp .&\pp {xxxxx}=
{\cal U}(S'S,\nu,p)
f_\pm(\nu,n,\pm (\Lambda'\Lambda)^{-1}p)^{{\it 0}\dots {\it 0}}
=U(S'S)f_\pm(\nu,n,\pm p)^{{\it 0}\dots {\it 0}}.
\ee
We have used here the fact that $\pi_{B}(S'S\nu,p)$ and
$\pi_{B}(S'\nu,p)$ are proportional (see (\ref{fp})) ---
technically this property of $\pi$-spinors is responsible for
the one-dimensionality of the representation and is typical only
of null momenta. 
An analog of (\ref{n-exD}) can be introduced also for the
massless fields. As an example consider again a field having
$n$-unprimed spinor indices. The corresponding index type of the
BW-spinor amplitude is represented by $n$ {\it 0\/}'s. 
Let us first trivially ``embed" the amplitude in the BW-spinor:
\be
f_\pm(\nu,n,\pm p)^{{\cal A}_1\dots {\cal A}_n}=
\left(
\begin{array}{c}
f_\pm(\nu,n,\pm p)^{{\it 0}\dots {\it 0}}\\
0\\
\vdots\\
0
\end{array}
\right).\label{embed}
\ee
Now  
\be
\parallel \psi_{\pm\alpha_1\dots\alpha_n}\parallel'^2&=&
\int d\mu_0(p)
f_\pm(\pm p)^{{\cal A}_1\dots {\cal A}_n}
\bar f_\pm(\pm p)^{{\cal B}_1\dots {\cal B}_n}
\vs_{{\cal A}_1{\cal B}_1}\dots
\vs_{{\cal A}_n{\cal B}_n}\nonumber\\
&=&
\int d\mu_0(p)
|f_\pm(\nu,n,\pm p)^{{\it 0}\dots {\it 0}}|^2
\label{n-0}.
\ee
The embedding (\ref{embed})
unifies the massive and massless cases because the spinor  
(\ref{embed}) is a true $SU(2)$ BW-spinor as opposed to the
amplitude $f_\pm(\nu,n,\pm p)^{{\it 0}\dots {\it 0}}$
which, taken alone, should be regarded as a $U(1)$
field.

\section{Summary and discussion}

The well known BW scalar product is a particular
case of a larger class of scalar products parametrized by a
family of world-vectors. If the world-vectors are null and
$p$-dependent then the BW amplitudes play a role of
momentum-space wave functions corresponding to projections 
of the Pauli-Lubanski vector in these momentum-dependent null
directions. The choice of null directions leads to a
simplification of the formalism because of the factorization
property of null world-vectors. The BW amplitudes constructed
in this way transform as scalar fields under the action of
active $SL(2,C)$ transformations. The corresponding passive
transformations of the amplitudes are local (i.e. $p$-dependent)
$SU(2)$. 
The BW indices, which originally played a
role of binary numbering of different irreducible components
of the spinor BW field,
turn out to play a dual role of BW-spinor indices if the passive
transformations are concerned. This property leads to a
BW-analog of the ordinary spinor algebra. 

There exist also other interesting formal
 analogies between the BW-spinor and
2-spinor formalisms. For example, the 
BW-spinors are obtained as contractions of 2-spinor indices of BW
fields with $\omega$-spinors. This is analogous to the way
Penrose {\it et al.\/} introduce spin-weighted spherical
harmonics \cite{PR} but here everything happens in the
momentum-space. It seems that the similarities between the two
approaches are worth of further studies. This includes the 
question of the role of conformal symmetries (typical of {\it
null\/} formalisms) and relations to twistors. 

The fact that pairs of null directions corresponding to
spin-frames simplify the description of
infinite-dimensional unitary representations of the Poincar\'e
group
goes hand-in-hand with a general philosophy underlying the spinor
approach to field theories and space-time geometry. 
In a forthcoming paper we shall discuss implications of the
generalized formalism for the structure of generators. One may
expect that the null formalism will be related to Dirac's front
form of generators \cite{dirac}.

\section{Acknowledgement}

Main results of this paper were obtained during my stay in
Oaxtepec, Mexico. I would like to thank Prof. Bogdan Mielnik for
the invitation, hospitality, and comments on this work.

\section{Appendix: Bispinors vs. BW-spinors}

An unprimed lower-index bispinor is
\be
\psi_\alpha=
\left(
\begin{array}{c}
\psi^{\it 0}_A\\
\psi^{\it 1}_{A'}
\end{array}
\right). 
\ee
Complex conjugated bispinors are 
\be
\bar \psi_{\alpha'}=\overline{\psi_\alpha}=
\left(
\begin{array}{c}
\overline{\psi^{\it 0}_A}\\
\overline{\psi^{\it 1}_{A'}}
\end{array}
\right)=
\left(
\begin{array}{c}
\bar \psi^{\it 1}_{A'}\\
\bar \psi^{\it 0}_{A}
\end{array}
\right).
\ee
Let
\be
\psi^{\it 0}_\alpha=
\left(
\begin{array}{c}
\psi^{\it 0}_A\\
0
\end{array}
\right),\quad\quad
\psi^{\it 1}_\alpha=
\left(
\begin{array}{c}
0\\
\psi^{\it 1}_{A'}
\end{array}
\right),\quad\quad
\bar \psi^{\it 1}_{\alpha'}=
\left(
\begin{array}{c}
\bar \psi^{\it 1}_{A'}\\
0
\end{array}
\right),\quad\quad
\bar \psi^{\it 0}_{\alpha'}=
\left(
\begin{array}{c}
0\\
\bar \psi^{\it 0}_{A}
\end{array}
\right),
\ee
\be
\omega_\alpha=
\left(
\begin{array}{c}
\omega_A\\
\bar \omega_{A'}
\end{array}
\right),\quad\quad
\bar \omega_{\alpha'}=
\left(
\begin{array}{c}
\bar \omega_{A'}\\
\omega_{A}
\end{array}
\right).
\ee
Then 
\be
\omega^\alpha\psi^{\it 0}_\alpha&=&
\omega^A\psi^{\it 0}_A=\psi^{\it 0},\\
\omega^\alpha\psi^{\it 1}_\alpha&=&
\bar \omega^{A'}\psi^{\it 1}_{A'}=\psi^{\it 1},\\
\bar \omega^{\alpha'}\bar \psi^{\it 1}_{\alpha'}&=&
\bar \omega^{A'}\bar \psi^{\it 1}_{A'}=\bar \psi^{\it 1},\\
\bar \omega^{\alpha'}\bar \psi^{\it 0}_{\alpha'}&=&
\omega^{A}\psi^{\it 0}_{A}=\bar \psi^{\it 0}.
\ee
The following expression appears often in connection with 
BW scalar products:
\be
\omega^{A}\bar \omega^{A'}
\Bigl(
\psi^{\it 0}_A\overline{\psi^{\it 0}_A}
+
\psi^{\it 1}_{A'}\overline{\psi^{\it 1}_{A'}}
\Bigr) 
&=&
\omega^{A}\bar \omega^{A'}\Bigl(
\psi^{\it 0}_A\bar \psi^{\it 1}_{A'}
+
\psi^{\it 1}_{A'}\bar \psi^{\it 0}_{A}
\Bigr) 
=
\omega^{\alpha}\bar \omega^{\alpha'}
\Bigl(
\psi^{\it 0}_{\alpha}\bar \psi^{\it 1}_{{\alpha}'}
+
\psi^{\it 1}_{{\alpha}'}\bar \psi^{\it 0}_{{\alpha}}
\Bigr)\nonumber\\
&=&
\omega^{\alpha}\bar \omega^{\alpha'}
\psi^{\cal A}_{\alpha}\bar \psi^{\cal B}_{{\alpha}'}
\varsigma_{\cal AB}=
\psi^{\cal A}\bar \psi^{\cal B}
\varsigma_{\cal AB},
\ee
where 
\be
\varsigma_{\cal AB}=
\left(
\begin{array}{cc}
0 & 1 \\
1 & 0 
\end{array}
\right)=-\varsigma^{\cal AB}.\label{vs}
\ee
To express covariantly unimodularity of the BW transformation
matrices we introduce the BW-spinor version of $\ve$-spinors:
\be
\ve_{\cal AB}=
\left(
\begin{array}{cc}
0 & 1 \\
-1 & 0 
\end{array}
\right)=\ve^{\cal AB}.\label{ve}
\ee
(\ref{ve}) are used to raise or lower the BW-spinor indices.



\begin{references} 
\bibitem[*]{*}Electronic address: mczachor@sunrise.pg.gda.pl
\bibitem{I}M.~Czachor, Manifestly Covariant 
Approach to Bargmann-Wigner Fields (I):
Generalized scalar products and Wigner states, submitted to
Proc. Roy. Soc. London. 
\bibitem{BW}V.~Bargmann and E.~P.~Wigner, 
Group theoretical discussion of relativistic wave
equations, 
Proc. Nat. Acad. Sci.
USA {\bf 34}, 211 (1948).
\bibitem{PR}R.~Penrose and W.~Rindler, {\it Spinors and
Space-Time\/}, vol.~1 (Cambridge University Press, 1984).
\bibitem{Ohnuki}Y.~Ohnuki, {\it Unitary Representations of the
Poincar\'e Group and Relativistic Wave Equations\/} (World
Scientific, Singapore, 1988).
\bibitem{dirac}P.~A.~M.~Dirac, 
Forms of relativistic dynamics, {\it Rev. Mod. Phys.\/} {\bf
21}, 392 (1949).
\end{references}
\end{document}